\documentclass[12pt,a4paper,dvips]{article}

\usepackage{amssymb,amsmath,graphicx,a4p}
\usepackage{cite,mcite}

\usepackage{ifthen}
\usepackage{l3_title}
\usepackage{Lep}

\usepackage{higgs_input}
\journalname{Phys. Lett. B}
\date{8 June 1999}
\preprint{99-080}
\Lep{2}
\usepackage[nolists,noheads,nomarkers,tablesfirst]{endfloat}
\begin{document}
\begin{titlepage}
  
  \title{Search for the Standard Model Higgs boson \\
    in \boldmath{\epem} interactions at $\boldsymbol{\rts} = 189 \GeV$}
  
  \author{L3 Collaboration}
%
% The abstract
%
  \begin{abstract}
    A search for the Standard Model Higgs boson is carried out on
    176.4\pb of data collected by the L3 detector at a center-of-mass
    energy of 189\GeV.  The data are consistent with the expectations
    of Standard Model processes and no evidence of a Higgs signal is
    observed.  Combining the results of this search with those at
    lower center-of-mass energies, a lower limit on the mass of the
    Standard Model Higgs boson of 95.3\GeV is set at the 95\%
    confidence level.
  \end{abstract}
  
%
% Adds "To be submitted to ..." or "Submitted to ...", if relevant
%
\submitted
%\vfill
%\texttt{/afs/cern.ch/user/a/aaron/public/papers/SM\_1999/SM\_Higgs\_1999\_v4.3.ps.gz}
\end{titlepage}
%
%%%%%%%%%%%%%%%%%%%%%%%%%%%%%%%%%%%%%%%%%%%%%%%%%%%%%%%%%%%%%%%%%%%%%%%%%%%%%%%
% Introduction
%%%%%%%%%%%%%%%%%%%%%%%%%%%%%%%%%%%%%%%%%%%%%%%%%%%%%%%%%%%%%%%%%%%%%%%%%%%%%%%
%
\section{Introduction}
\label{sec:intro}
In the Standard Model~\cite{sm_glashow}, a
single Higgs doublet~\cite{higgs_1} gives rise to a
neutral scalar, the Higgs boson, with a mass, \mH, that is a free
parameter of the theory.  Searches in \epem collisions for the
Standard Model Higgs boson have been reported up to center-of-mass
energies of 183\GeV by L3~\cite{l3_1998_11} and other
experiments~\cite{opal_4}.  No evidence of a
signal has been found and a combined lower limit of
89.7\GeV~\cite{LEPHWG_183} is set at the 95\% confidence level.  In
this letter, the results of a Higgs search performed on the data
sample collected by L3 at $\rts=189\GeV$ are reported, significantly
extending the accessible range of \mH.

The dominant Higgs production mode,
\begin{displaymath}
  \epemtoSMHZ \; ,
\end{displaymath}
as well as the smaller production processes of \WW and \ZZ fusion,
are considered.
% in the \SMHZtobbnn and \SMHZtobbee channels, respectively.  
All significant signal decay modes are considered in the search.
Four-fermion final states from \W- and \Z-pair production, as well as
\epemtoqq, make up the largest sources of background.

%%%%%%%%%%%%%%%%%%%%%%%%%%%%%%%%%%%%%%%%%%%%%%%%%%%%%%%%%%%%%%%%%%%%%%%%%%%%%%%
% Data/MC
%%%%%%%%%%%%%%%%%%%%%%%%%%%%%%%%%%%%%%%%%%%%%%%%%%%%%%%%%%%%%%%%%%%%%%%%%%%%%%%
%
\section{Data and Monte Carlo samples}
\label{sec:datamc}
The data were collected using the L3
detector~\cite{l3_1990_1}
at LEP during 1998.  The integrated luminosity is 176.4\pb at an
average center-of-mass energy of 188.6\GeV.

Higgs production cross sections and branching ratios are calculated
using the HZHA generator~\cite{lepii_v2}, whereas for the efficiency
studies, Monte Carlo samples of Higgs events are generated using
PYTHIA~\cite{pythia_1}.  Standard Model background estimates
are made with the following Monte Carlo programs: PYTHIA (\epemtoqqg),
KORALW~\cite{koralw_1} (\epemtoWW), KORALZ~\cite{koralz}
(\epemtotautau), PYTHIA and PHOJET~\cite{phojet_1}
(\epemtoeeqq), EXCALIBUR~\cite{excalibur} (\epemtoffff) and PYTHIA and
EXCALIBUR ($\epem\!\rightarrow\!\Z\epem$).  The number of simulated
events for the most important background channels is at least 100
times the number of collected data events for such processes, while
the number of signal events is at least 300 times the number expected
to be observed in the data with this integrated luminosity.

The response of the L3 detector is simulated using the GEANT~3.15
program~\cite{geant}, taking into account the effects of multiple
scattering, energy loss and showering in the detector. Hadronic
interactions in the detector are modeled using the GHEISHA
program~\cite{gheisha}.

%%%%%%%%%%%%%%%%%%%%%%%%%%%%%%%%%%%%%%%%%%%%%%%%%%%%%%%%%%%%%%%%%%%%%%%%%%%%%%%
% Analysis section
%%%%%%%%%%%%%%%%%%%%%%%%%%%%%%%%%%%%%%%%%%%%%%%%%%%%%%%%%%%%%%%%%%%%%%%%%%%%%%%
%
\section{Analysis procedures}
The search procedure is dictated by the four event topologies
representing approximately 98\% of the \SMHZ decay modes:
$\qqbar\qqbar$, $\qqbar\nnbar$,
$\qqbar\ll\;(\ell=\mathrm{e},\mu,\tau)$ and $\tautau\qqbar$.  With the
exception of \SMHZtottqq, the analyses for each channel are optimized
for $\bigH\!\rightarrow\!\bbbar$, since this represents about 85\% of
the Higgs branching fraction in the mass range of interest. However,
the efficiencies for the smaller contributions from
$\bigH\!\rightarrow\!\ccbar,\mathrm{gg}$ are also considered.

The analyses for all the channels are performed in three stages.
First, a high multiplicity hadronic event selection is applied,
greatly reducing the large background from two-photon processes, while
at the same time maintaining a high efficiency for the Higgs signal
over a broad range of masses.  Second, a tighter set of cuts specific
to the topology in question is used to further enrich the sample of
events while still maintaining signal efficiencies on the order of
50\%.  Finally, a discriminating variable is built for each analysis.
These discriminants include the results of a kinematic fit of the
event, imposing 4-momentum conservation, and depend on the mass
hypothesis value, \mH.  The spectra of the discriminants are computed for
the observed data and the Monte Carlo backgrounds and signals at each value
of \mH considered, in the range $50\GeV\leq\mH\leq 100\GeV$.

The b-tagging variable, used to identify b quarks, plays a major role
in the computation of the final discriminant.  A neural
network~\cite{aaron_thesis,l3_1997_18} is used to calculate the b-tag
for each hadronic jet from the three-dimensional decay lengths,
semileptonic information and jet-shape variables.  The b-tag variable
used for the entire event is a combination of the individual jet-tag
probabilities.

%%%%%%%%%%%%%%%%%%%%%%%%%%%%%%%%%%%%%%%%%%%%%%%%%%%%%%%%%%%%%%%%%%%%%%%%%%%%%%%
% Hqq analysis
%%%%%%%%%%%%%%%%%%%%%%%%%%%%%%%%%%%%%%%%%%%%%%%%%%%%%%%%%%%%%%%%%%%%%%%%%%%%%%%
%
\subsection{The \SMHZtobbqq channel}
Events from this channel usually consist of four jets, two of which
contain b hadrons, while the other two have a mass consistent with the
\Z mass.  Standard Model processes which mimic these events are
typically four-jet final states from \qqbar with hard gluons, \WW and
\ZZ events, especially those where one of the \Z bosons decayed into b
quarks.

Two independent analyses of this channel are performed, a neural
network approach and a cut-based likelihood analysis.  The cut-based
analysis is the primary one used in this channel.  The neural network
analysis achieves similar performance and a description of the
analysis technique can be found in Reference~\cite{l3_1998_11}.

First, a preselection designed to accept high multiplicity hadronic
events is applied by requiring at least 15 charged tracks and 20
calorimetric clusters.  The visible energy, \Evis, must be between
$0.6\rts$ and $1.4\rts$.  The missing energy parallel and
perpendicular to the beam direction has to be less than $0.3\Evis$.
Finally, the energy of the most energetic photon or lepton must be
less than 65\GeV.

At this stage, all the events passing the preselection are forced to
have four jets using the DURHAM~\cite{DURHAM} clustering algorithm and
a kinematic fit requiring 4-momentum conservation is performed.  An
automated procedure~\cite{l3_1998_16,aaron_thesis} is used to optimize
the selection criteria, which differentiates the Higgs signal from
background based mainly on kinematic differences and large b-tag
values.  The cuts chosen by the optimizer, which do not depend on the
\mH hypothesis, are as follows.  To reject gluonic jets in \qqbar
events, the dijet masses must be between $0.13\rts$ and $0.63\rts$;
the minimum jet energy must be larger than $0.14\rts$ and the maximum
energy difference between any two jets must be less than $0.22\rts$.
To enhance the four-jet nature of the events, the $Y_{\mathrm{cut}}$
parameter in the DURHAM scheme where the event goes from a three-jet
to a four-jet topology, \Ytf, is required to be larger than 0.0086.
Finally, there must be at least 22 charged tracks.  As in previous
publications~\cite{l3_1998_16}, the $\chi^2$-probability that depends
on \mH and \mZ is used to quantify the consistency of the event with a
given \mH hypothesis.  A loose cut is placed on this variable, but
more importantly, it is used along with the b-tag to calculate the
mass-dependent final discriminant.

\begin{figure}[htbp]
  \begin{center}
    \includegraphics*[width=\textwidth, bb=0 220 670 981]{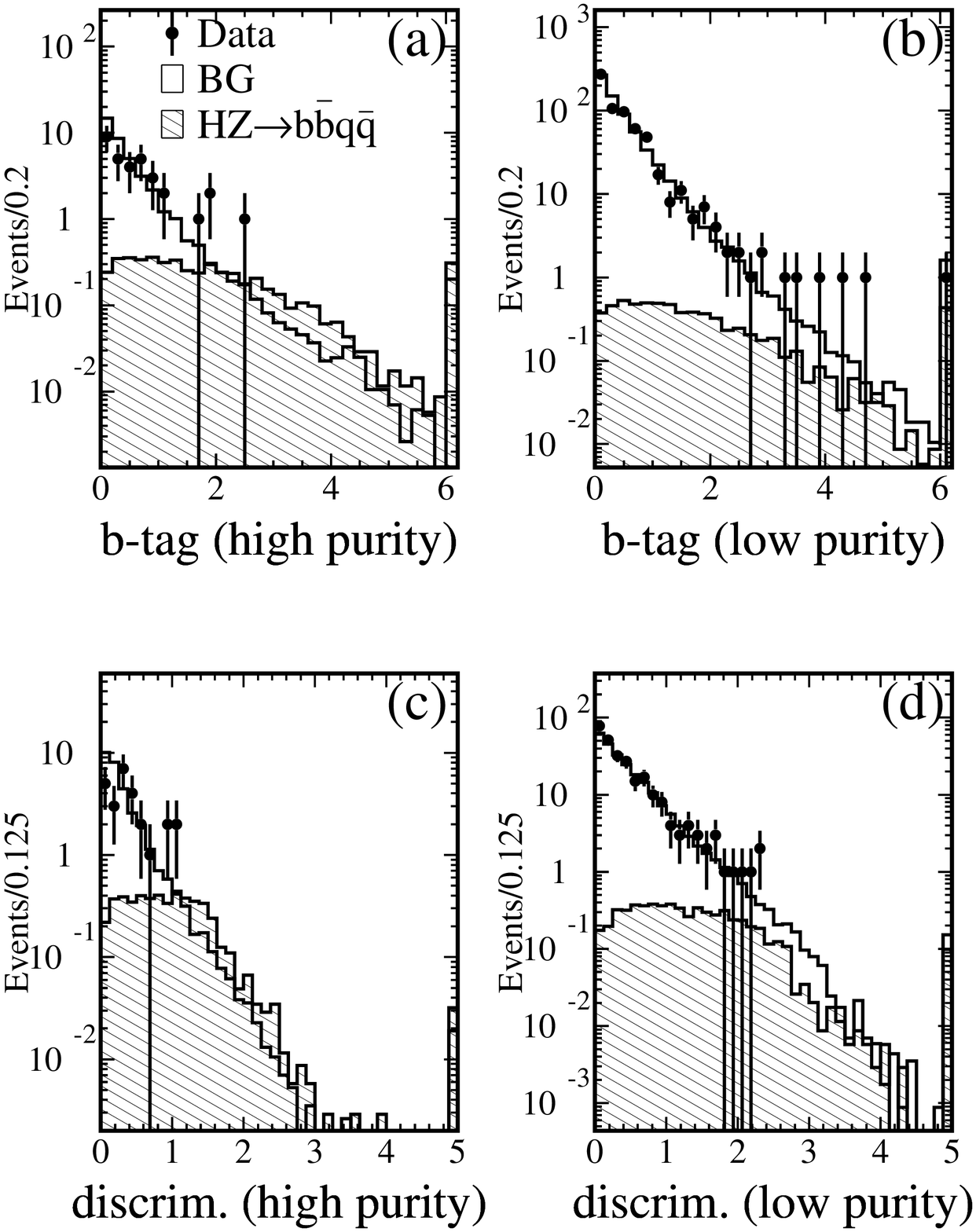} 
    \caption{The b-tag distributions for the (a) high purity and (b) low purity
      analyses, and the final discriminant for the (c) high purity and
      (d) low purity analyses of \SMHZtobbqq.  The points are the
      189\GeV data, the open histograms are Monte Carlo background and
      the hatched histograms are the Higgs signal.  The signal is
      shown for \SMHZtobbqq with $\mH=95\GeV$ and is normalized to the
      number of expected events. The last bin in each histogram
      contains the overflows.}
    \label{fig:hqq_btag_final}
  \end{center}
\end{figure}

At this point, 682 events remain in the data and 703 in the Monte
Carlo background, with 85\% of these from \WW events.  These four-jet
\WW events are characterized by their low b-tag values and the
consistency of the dijet masses with \mW.  With this in mind, the
optimizer splits the surviving events into high purity and low purity
samples using a sliding cut on the reconstructed dijet mass, \Meq,
from a kinematic fit assuming the five constraints (5C) of 4-momentum
conservation and equal masses for the two dijet systems.  If $\Meq >
0.74 \mH + 21.7 \GeV$, then the event is placed into the high purity
category, otherwise it is placed into the low purity category.  The
low purity sample contains most of the properly reconstructed \WW
events, isolating a large component of the dominant background.  In
the high purity sample, the optimizer chooses b-tag values to be
larger than 0.09, while in the low purity sample the cut is 0.41.  The
b-tag spectra for these two samples are shown in
Figures~\ref{fig:hqq_btag_final}(a) and~\ref{fig:hqq_btag_final}(b).

After the final set of cuts, the high purity category contains 26
candidates with 31.0 expected from Standard Model processes and a
signal efficiency of 28\% for \SMHZtobbqq with $\mH=95\GeV$.  The low
purity category has 263 candidates, 239 expected background events
and a 34\% signal efficiency.

Once the final set of cuts has been applied, the weighted
probability~\cite{l3_1998_16,aaron_thesis} that an event is
consistent with the background distributions of both the b-tag and the
mass variable is calculated.  Since the weighted combination depends
on the mass hypothesis, \mH, the distributions of this discriminant
are calculated for each test mass, an example of which can be found in
Figures~\ref{fig:hqq_btag_final}(c) and~\ref{fig:hqq_btag_final}(d).  The
observed candidates in the \SMHZtobbqq channel are consistent with the
Monte Carlo background predictions.

%%%%%%%%%%%%%%%%%%%%%%%%%%%%%%%%%%%%%%%%%%%%%%%%%%%%%%%%%%%%%%%%%%%%%%%%%%%%%%%
% Hnn Analysis
%%%%%%%%%%%%%%%%%%%%%%%%%%%%%%%%%%%%%%%%%%%%%%%%%%%%%%%%%%%%%%%%%%%%%%%%%%%%%%%
%
\subsection{The \SMHZtobbnn channel}
This channel is characterized by two acoplanar jets and large missing
transverse energy.  The missing mass is consistent with \mZ and the
hadronic jets typically contain b hadrons.

Two independent analyses of this channel are carried out, a neural
network and a cut-based likelihood analysis.  The analyses have
similar performance and lead to consistent results.  In this letter,
the neural network analysis is described.

First, high multiplicity hadronic events with more than 3 charged
tracks and at least 15 calorimetric energy clusters are selected.
Using the DURHAM algorithm, all energy clusters in the event are
combined to form two hadronic jets. The reconstructed mass of each of
these jets must exceed 40\GeV.  These cuts reduce contributions from
purely leptonic two-fermion final states, as well as two-photon
interactions, while keeping a significant fraction of hadronic events
from \epemtoqqg and \W-pair production.  These latter contributions
are further reduced by requiring the visible mass to be less than
120\GeV and the mass recoiling against the hadronic system to lie
between 50\GeV and 130\GeV.

\begin{figure}[htbp]
  \begin{center}
    \includegraphics*[width=\textwidth,bb=0 220 670 981]{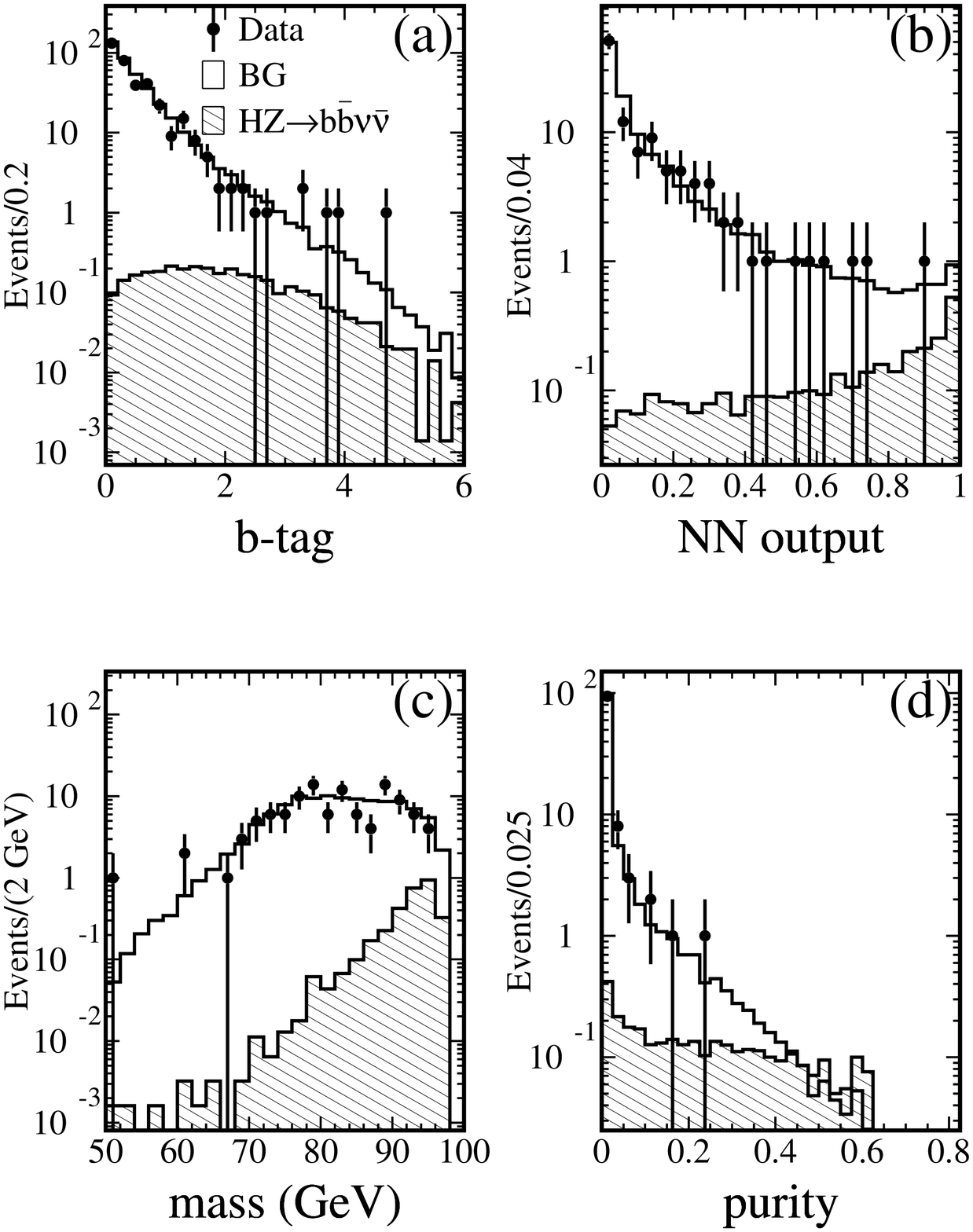} 
    \caption{Distributions of the (a) b-tag, (b) neural network output,
      (c) hadronic mass, \MfiveC, and (d) purity variable for the
      \SMHZtobbnn analysis.  The points are the 189\GeV data, the open
      histograms the background and the hatched histograms are for
      \SMHZtobbnn with $\mH=95\GeV$, normalized to the number of
      expected events.}
    \label{fig:hnn_btag_nnet_mass_final}
  \end{center}
\end{figure}

Events from \epemtoqqg are further suppressed with missing-energy
requirements.  The missing energy transverse to the beam axis should
be greater than 5\GeV, the missing momentum vector must be at least
$16^\circ$ from this axis and the longitudinal missing energy is
required to be less than $0.7\rts$.  The opening angle between the two
jets has to be greater than $69^\circ$ and the angle between the
jet-jet plane and the beam-axis must be greater than $3^\circ$. The
energy in the forward luminosity calorimeter is required to be smaller
than 15\GeV.  In addition, the event b-tag must be larger than 0.5.
The b-tag spectra for data and Monte Carlo are shown in
Figure~\ref{fig:hnn_btag_nnet_mass_final}(a). After this final set of
cuts, there remain 109 data events, with 116 expected from Standard
Model processes and an efficiency of 62\% for \SMHZtobbnn with
$\mH=95\GeV$.

A mass-independent neural network~\cite{l3_1998_11} is then used to
further separate the signal from background.  A kinematic fit imposing
4-momentum conservation and requiring the missing mass to be \mZ is
performed, yielding the hadronic mass, \MfiveC.  The neural network
output is shown in Figure~\ref{fig:hnn_btag_nnet_mass_final}(b), and
the distribution of \MfiveC is shown in
Figure~\ref{fig:hnn_btag_nnet_mass_final}(c).  The \MfiveC mass is
combined with the neural network output to form the purity
variable~\cite{l3_1998_11}.  This purity variable plays the role of
the final discriminant in the \SMHZtobbnn analysis and is shown in
Figure~\ref{fig:hnn_btag_nnet_mass_final}(d) for the mass hypothesis
$\mH=95\GeV$.  The observed data in the \SMHZtobbnn analysis are
compatible with the Monte Carlo background expectations.

%%%%%%%%%%%%%%%%%%%%%%%%%%%%%%%%%%%%%%%%%%%%%%%%%%%%%%%%%%%%%%%%%%%%%%%%%%%%%%%
% Hee Hmm Analysis
%%%%%%%%%%%%%%%%%%%%%%%%%%%%%%%%%%%%%%%%%%%%%%%%%%%%%%%%%%%%%%%%%%%%%%%%%%%%%%%
%
\subsection{The \SMHZtobbee and \SMHZtobbmm channels}
The signatures for \SMHZtobbee and \SMHZtobbmm are a pair of high
energy electrons or muons, with an invariant mass near \mZ, and two
hadronic b jets.

A hadronic event selection is applied requiring at least 5 charged
tracks, 15 calorimetric clusters and two well identified electrons or
muons.  The visible energy must be larger than $0.7\rts$ for the
electron analysis and $0.4\rts$ for the muon analysis.  In the
\SMHZtobbee analysis, the electron pair must have an opening angle
greater than $100^\circ$, while for \SMHZtobbmm, the muon pair must
have an opening angle greater than $90^\circ$. In addition, there must
be less than $0.4\rts$ of missing energy perpendicular to the beam
direction.  Both analyses must have values of \Ytf larger than 0.0009.
Finally, the invariant mass of the leptonic system after a kinematic
fit imposing 4-momentum conservation must be between 60\GeV and
110\GeV for the electrons and 50\GeV and 125\GeV for the muons.  After
this final set of cuts, the number of remaining candidates in the
electron channel is 15, with 13.2 expected from Standard Model
backgrounds and a signal efficiency for \SMHZtobbee of 77\% for
$\mH=95\GeV$.  The corresponding numbers for the muon channel are 5
candidates with 5.5 background expected and a signal efficiency for
\SMHZtobbmm of 57\%.

\begin{figure}[htbp]
  \begin{center}
    \includegraphics*[width=\textwidth, bb=0 220 670 981]{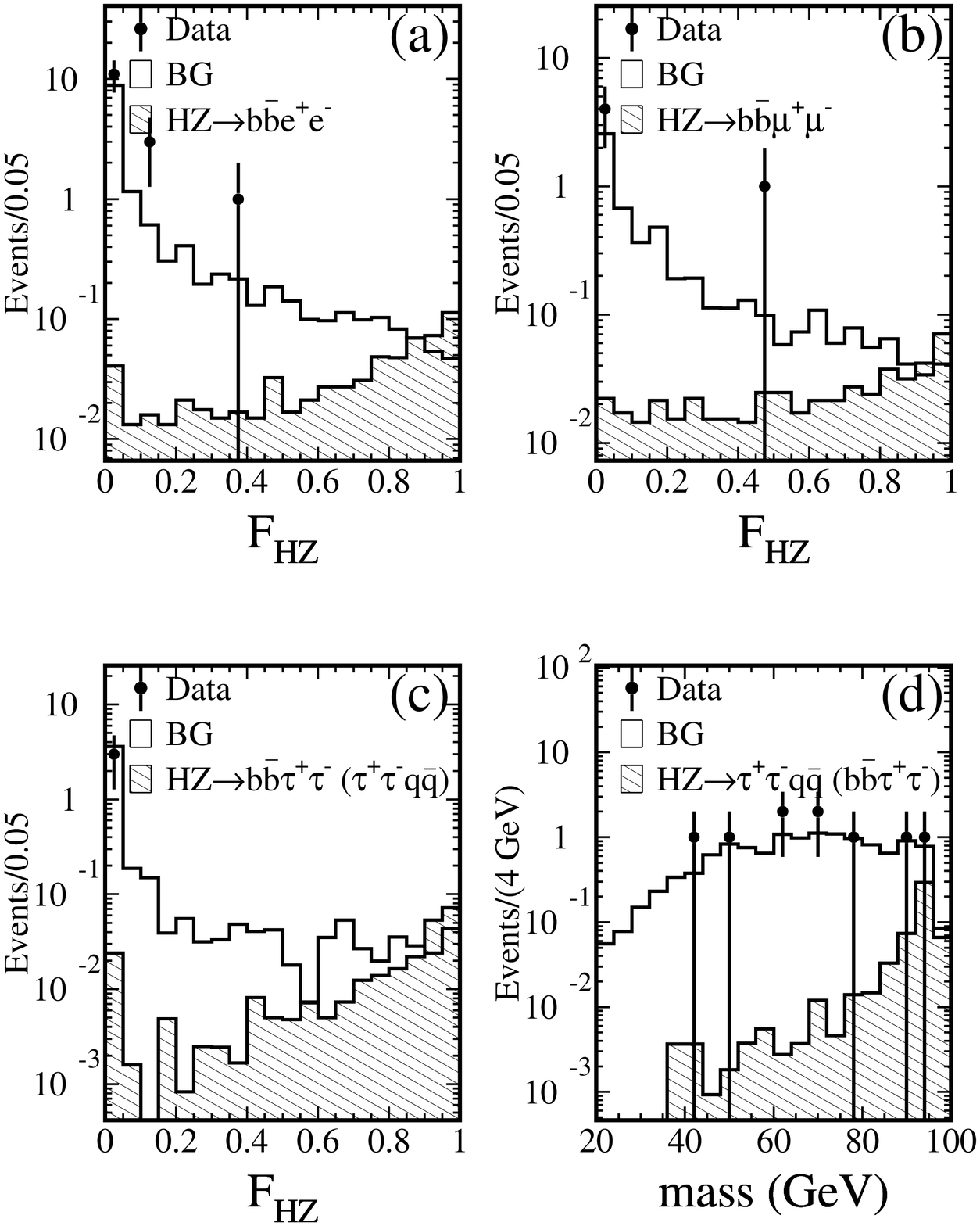} 
    \caption{Distributions of the final discriminant for the (a) \SMHZtobbee,
      (b) \SMHZtobbmm, (c) \SMHZtobbtt and (d) \SMHZtottqq channels
      for the 189\GeV data, background and a Higgs signal of 95\GeV,
      normalized to the number of expected events. The signal events
      in the \SMHZtobbtt and \SMHZtottqq plots include the
      branching-ratio-corrected cross-efficiencies for these channels.
      Events are uniquely assigned to only one of these channels.}
    \label{fig:hll_final}
  \end{center}
\end{figure}

After performing a kinematic fit requiring 4-momentum conservation and
constraining the mass of the lepton pair to \mZ, the mass of the
jet-jet system is combined with the b-tags of jet 1 and jet 2.  For
each event class $j$ (\ZZ,\;\WW,\; \qqbar,\;$\Z\epem$,\;\SMHZ), a
probability density function $f_j^i$ is constructed, where $i$ denotes
the b-tag of jet 1, the b-tag of jet 2, or the dijet mass.  The
probability of an event to belong to class $j$, based solely on the
value of the variable $i$, is then defined as
\begin{equation}
  \label{eq:hll_probability}
  p_j^i=\frac{f_j^i}{\sum_k f_k^i} \; .
\end{equation} 
Finally, the probabilities for the individual variables are combined
by calculating the likelihood that the event belongs to the signal
class \SMHZ:
\begin{equation}
  \label{eq:hll_likelihood}
  F_{\SMHZ}=\frac{\prod_i p_{\SMHZ}^i}{\sum_k\prod_i p_k^i} \; .
\end{equation}
The spectra for this final discriminant, $F_\SMHZ$, in the electron
and muon channels are shown in Figures~\ref{fig:hll_final}(a)
and~\ref{fig:hll_final}(b) for the data, background and a 95\GeV Higgs
signal.  The observed candidates are consistent with the Monte Carlo
background predictions.

%%%%%%%%%%%%%%%%%%%%%%%%%%%%%%%%%%%%%%%%%%%%%%%%%%%%%%%%%%%%%%%%%%%%%%%%%%%%%%%
% Hll and ttqq Analysis
%%%%%%%%%%%%%%%%%%%%%%%%%%%%%%%%%%%%%%%%%%%%%%%%%%%%%%%%%%%%%%%%%%%%%%%%%%%%%%%
%
\subsection{The \SMHZtobbtt and \SMHZtottqq channels}
The \SMHZtobbtt and \SMHZtottqq final states are very similar and can be
distinguished only with mass and b-tag information.  The semileptonic
W- and Z-pair decays are the most significant background sources.

Two inclusive selections are performed, one based on a tau
identification (particle-based selection) and one relying more on the
event kinematics (jet-based selection). Events are accepted if they
pass either of the two selections.

First, a common preselection is applied, requiring more than 4 charged
tracks, more than 14 clusters and a visible energy of more than
$0.4\rts$.  The events are subject to the DURHAM algorithm, keeping
only those with \Ytf larger than 0.0025.  Background from
\epemtoqqg is reduced by rejecting events containing
photons with energies greater than 40\GeV.  The contribution of
$\WW\!\rightarrow\!\qqbar\ell\nu \; (\ell=\e,\mu)$ is reduced by
requiring the energy of electrons and muons to be smaller than 40\GeV.

In the particle-based selection, tau leptons are identified via their
decay into electrons or muons, or as an isolated low-multiplicity jet
with 1 or 3 tracks and unit charge.  In the jet-based selection, the
event is forced into four jets using the DURHAM algorithm. Two of the
jets must each have less than 4 tracks. These jets are considered as
tau candidates, but at least one of them must coincide with a tau
candidate defined in the particle-based selection within a $3^\circ$
cone. Both taus must be separated from the hadronic jets by at least
25$^\circ$. Background contamination from fully hadronic \WW decays is
reduced by rejecting events where both taus decay into 3 charged
particles and by requiring the visible energy to be smaller than
$0.95\rts$ for the particle-based and smaller than $0.9\rts$ for the
jet-based selection.  Moreover, in the jet-based selection, the polar
angle of the missing momentum vector, $\Theta_\mathrm{miss}$, must
satisfy $|\cos\Theta_\mathrm{miss}| \leq 0.95$ in order to reduce
$\qqbar(\gamma)$ contamination.

The invariant masses of the tau-tau and the jet-jet systems are
obtained from a kinematic fit which imposes 4-momentum conservation.
An event qualifies for the \SMHZtobbtt channel if the invariant mass
of the tau-tau system is consistent with the mass of the Z boson by
lying between 70\GeV and 125\GeV.  Similarly, an event qualifies for
the \SMHZtottqq channel if the jet-jet mass fulfills this same
requirement.  Furthermore, the opening angle of the particles or jets
assigned to the Higgs boson must be larger than $70^\circ$ and those
assigned to the \Z must be at least $100^\circ$ apart.
Cross-efficiencies on the \SMHZtobbee and the \SMHZtobbmm channels (up
to 3\%) are taken into account by rejecting events which were already
selected in those analyses.  In total, 12 candidate events are
selected, with 17.1 events expected from Standard Model background
processes passing either of the tau selections, and an efficiency of
30\% for both \SMHZtobbtt and \SMHZtottqq at $\mH=95\GeV$.

The final discriminant for the \SMHZtobbtt channel is defined
similarly to the likelihood used in the \SMHZtobbee and \SMHZtobbmm
analyses, using Equations~\ref{eq:hll_probability}
and~\ref{eq:hll_likelihood}.  For the \SMHZtottqq channel, the mass
distribution of the tau pair, after constraining the invariant mass of
the jets to \mZ, is used as the final discriminant.  Events that pass
both decay hypotheses are placed into the channel with the larger
value of the likelihood, $F_\SMHZ$, defined in
Equation~\ref{eq:hll_likelihood}.  Distributions of these
discriminants can be found in Figures~\ref{fig:hll_final}(c)
and~\ref{fig:hll_final}(d) for data, background and a 95\GeV Higgs
signal.  No evidence of a signal is seen in either of the tau
channels.

%%%%%%%%%%%%%%%%%%%%%%%%%%%%%%%%%%%%%%%%%%%%%%%%%%%%%%%%%%%%%%%%%%%%%%%%%%%%%%%
% The whole enchilada
%%%%%%%%%%%%%%%%%%%%%%%%%%%%%%%%%%%%%%%%%%%%%%%%%%%%%%%%%%%%%%%%%%%%%%%%%%%%%%%
%
\section{Combined results}
\label{sec:results}
\begin{table}[htbp]
  \begin{center}
    \caption{The number of expected signal (SIG), background (BG) events and
      observed candidates (DATA) for the $\rts=189\GeV$ data after a
      cut on the final discriminant corresponding to a
      signal-over-background ratio greater than 0.05.  The number of
      signal events includes cross-efficiencies from other signal
      channels, events from fusion processes, charm and gluonic Higgs
      decays.  Events are uniquely assigned to a channel.}\vspace{0.5em}
    \label{tab:results}
    \begin{tabular}{lllrrrrrrrrrrrr} \hline
      &&& \multicolumn{9}{c}{Mass hypothesis\rule{0em}{1.2em}} \\
      \multicolumn{2}{c}{Selection} && \multicolumn{3}{c}{$\mH=85\GeV$} && \multicolumn{3}{c}{$\mH=90\GeV$} && \multicolumn{3}{c}{$\mH=95\GeV$} & \\
      \cline{1-2} \cline{4-15} \\
      \bigH & \Z && SIG & BG & \multicolumn{2}{r}{DATA} & SIG & BG & \multicolumn{2}{r}{DATA} & SIG & BG & \multicolumn{2}{r}{DATA} \\
      \hline
% H       Z     | S85    B85     D85 | S90    B90    D90|  S95    B95    D95
\bbbar\rule{0em}{1.2em} 
       & \qqbar && 21.2 & 73.3 & 78 && 17.6 & 78.4 & 90 && 8.6 &  53.7 &  51 &\\
\bbbar & \nnbar && 9.0  & 21.3 & 22 && 5.9  & 16.0 & 17 && 2.3 &  9.2  &  4  &\\
\bbbar & \epem  && 1.8  & 3.0  & 3  && 1.2  & 3.5  & 1  && 0.6 &  2.3  &  1  &\\
\bbbar & \mumu  && 1.7  & 5.5  & 5  && 1.1  & 3.0  & 3  && 0.5 &  1.9  &  1  &\\
\bbbar & \tautau&& 0.7  & 1.2  & 0  && 0.4  & 0.8  & 0  && 0.2 &  0.5  &  0  &\\
\tautau& \qqbar && 1.6  & 5.0  & 3  && 1.1  & 4.0  & 3  && 0.4 &  2.4  &  2  &\\
      \hline
    \end{tabular}
  \end{center}
\end{table}

\begin{figure}[htbp]
  \begin{center}
    \includegraphics*[width=\textwidth, bb=0 220 670 981]{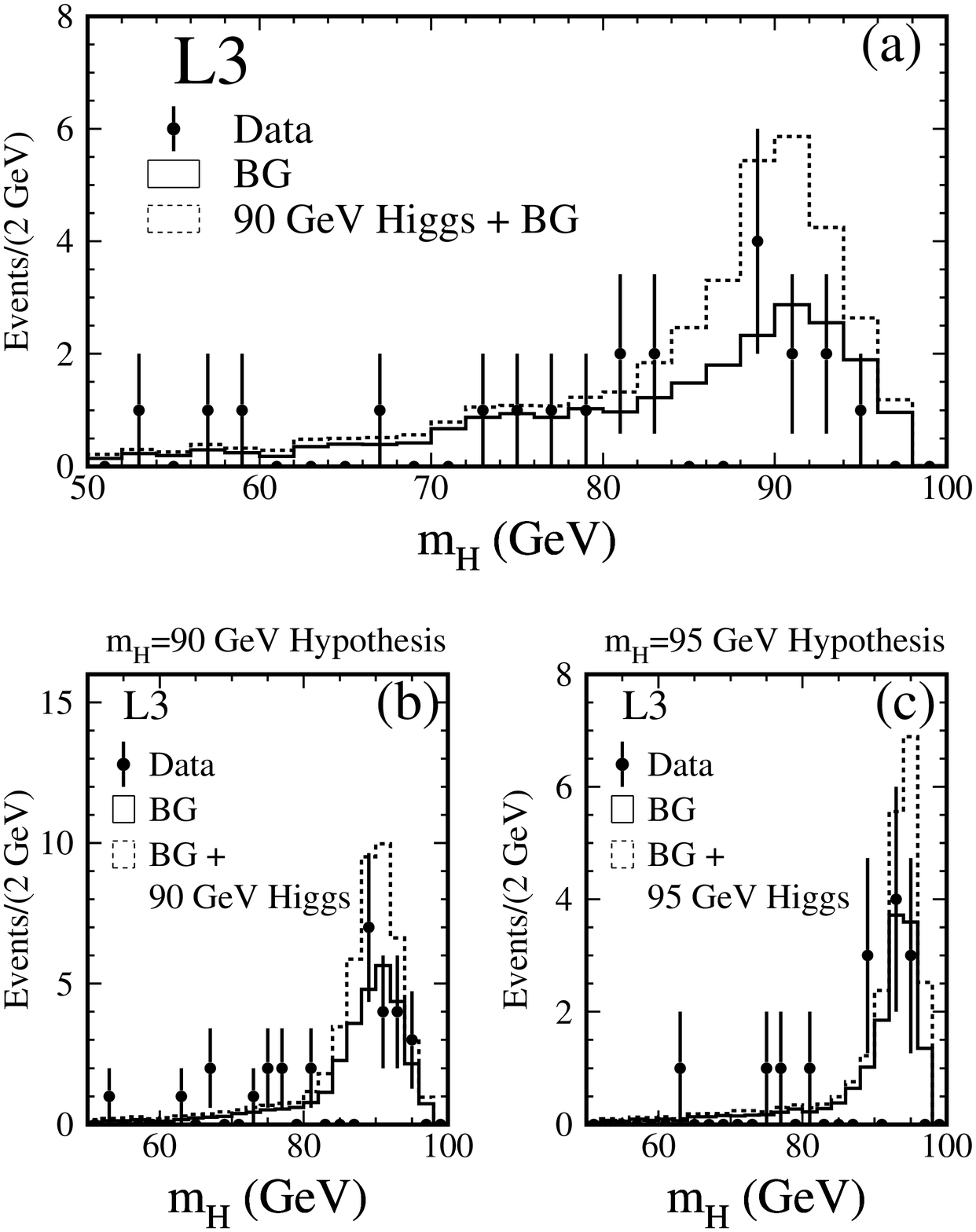} 
    \caption{Reconstructed Higgs mass distributions in the 189\GeV data for the most
      significant signal-like events of the various analyses: (a)
      after mass-independent cuts on b-tag or neural network output to
      select candidates; (b-c) after cuts on the final discriminant
      (signal-over-background ratio greater than 0.25) for the mass
      hypothesis (b) $\mH=90\GeV$ and (c) $\mH=95\GeV$.  In all plots,
      the points are the data, the solid histograms are the Monte
      Carlo background and the dashed histograms are the Monte Carlo
      background plus Higgs signal.}
    \label{fig:final_mass}
  \end{center}
\end{figure}

The results of all the previously described analyses are combined in
this section.  For illustrative purposes, in
Figure~\ref{fig:final_mass}(a) the reconstructed Higgs mass is shown
for a sample of signal-like events selected by the analyses after
making a mass-independent requirement, such as a large b-tag value or
large neural network output.  In Figures~\ref{fig:final_mass}(b)
and~\ref{fig:final_mass}(c), the results of the mass-dependent
selections are illustrated by plotting the reconstructed Higgs mass
for events with large discriminant values (signal-over-background
ratio greater than 0.25) for the 90\GeV and 95\GeV mass hypotheses.
No evidence of a signal is present in any of the analyses and a global
confidence level (CL) on the absence of a signal is calculated from
the spectra of final discriminants from all the analyses in a scan
over \mH from 50\GeV to 100\GeV.  The CL is calculated using the
techniques of References~\cite{l3_1997_18,new_method}, which also
allow correlated and statistical errors to be easily accounted for in
the computation of CL.

The statistical and systematic errors on the signal and background are
considered using the same procedure as previous Standard Model Higgs
searches by L3~\cite{l3_1998_11,l3_1997_18}.  The overall systematic
error is estimated at 10\% on the number of background events and 5\%
on the number of signal events.  The statistical error on the
background from the finite number of generated Monte Carlo events is
larger, but is uncorrelated from bin to bin in the final discriminant
distributions, and has little effect on the CL.  Bins with a
signal-over-background ratio smaller than 0.05 are not considered
during the calculation of CL.  This cut was chosen to maximize the
average CL, as calculated from a large number of Monte Carlo
experiments, thereby minimizing the degradation of the result due to
these systematic and statistical errors for this integrated luminosity
and center-of-mass energy.  The results of all the analyses after such
a signal-over-background cut are summarized in Table~\ref{tab:results}
for the data, Monte Carlo background and signal.  The number of signal
events includes cross-efficiencies from other channels, fusion
processes and charm and gluonic Higgs decays.

\begin{figure}[htbp]
  \begin{center}
    \includegraphics*[width=\textwidth, bb=0 220 670 981]{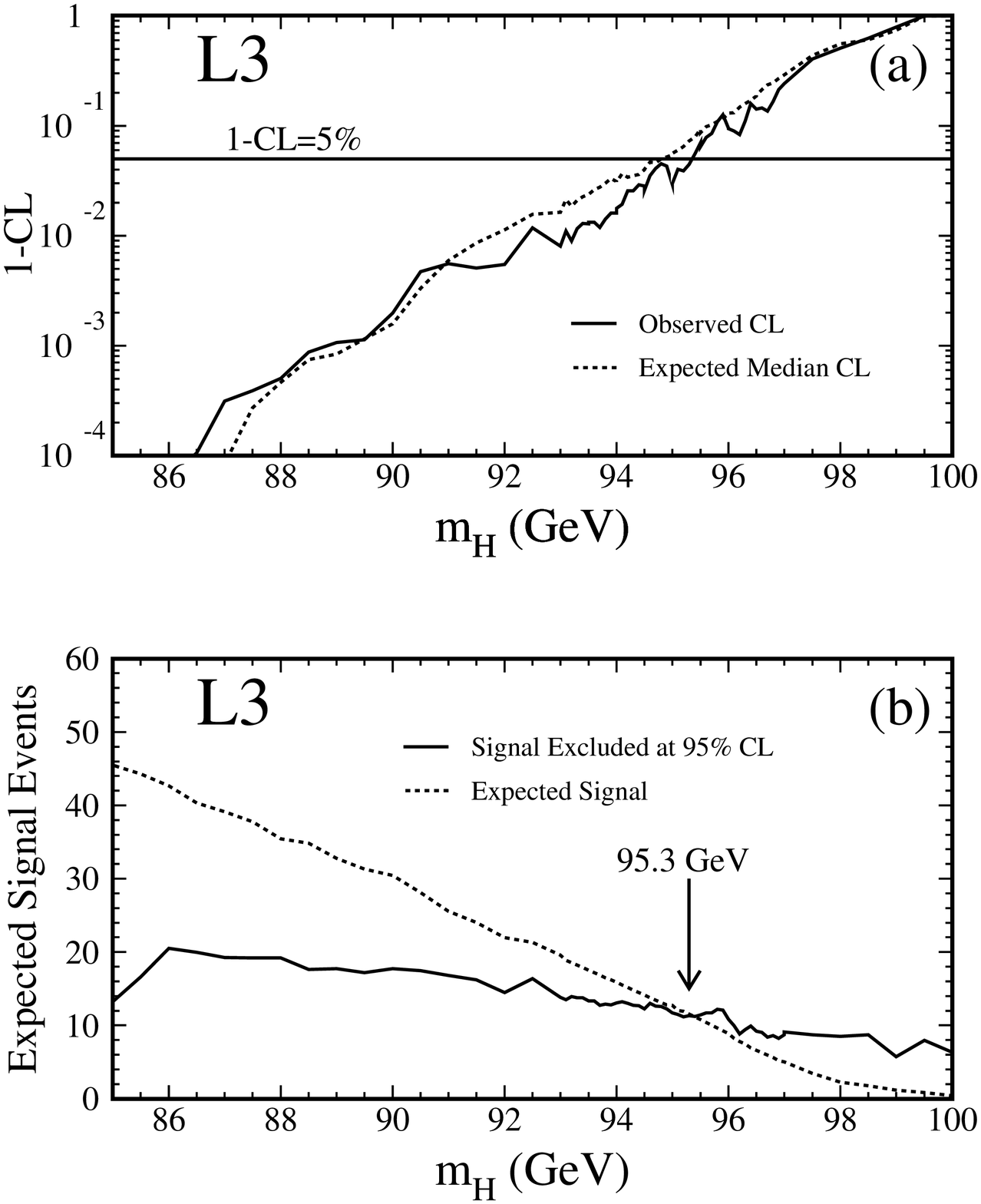} 
    \caption{(a) The observed and expected median confidence levels as
      a function of the Higgs mass. (b) The number of expected and
      excluded signal events. Both plots include results from lower
      center-of-mass energies.  The lower limit on the Higgs mass is
      set at $\mH>95.3\GeV$ at the 95\% CL.}
    \label{fig:limit}
  \end{center}
\end{figure}

The measured value of CL as a function of the Standard Model Higgs
boson mass, in the range $85\leq\mH\leq 100\GeV$, is shown in
Figure~\ref{fig:limit}(a), along with the median of the CL
distribution as calculated from a large sample of Monte Carlo
experiments assuming a background-only hypothesis. The number of Higgs
events expected to be observed, as a function of \mH, and the number
of excluded signal events at the 95\% CL are shown in
Figure~\ref{fig:limit}(b).  The results of previous L3 Standard Model
Higgs searches at lower center-of-mass
energies~\cite{l3_1998_11,l3_1997_18,l3_1996_12} have been included in
the calculation of these confidence levels.  Values of \mH from 50\GeV
to 85\GeV are excluded to greater than the 99.999\% confidence level
by the 189\GeV data alone and have been previously excluded by the L3
analyses at lower center-of-mass energies.  The median CL represents
the sensitivity of the global analysis and is equal to 95\% at
$\mH=94.8\GeV$, while the average CL is 95\% at 92.7\GeV.  Where the
observed CL falls below 95\%, the probability to observe a higher
limit is 37\%.

The lower limit on the Standard Model Higgs boson mass is set at
\begin{displaymath}
  \mH > 95.3\GeV \; \mathrm{at\ 95\% \ CL.}
\end{displaymath}
This new lower limit improves upon and supersedes our previously
published results.


%
%%%%%%%%%%%%%%%%%%%%%%%%%%%%%%%%%%%%%%%%%%%%%%%%%%%%%%%%%%%%%%%%%%%%%%%%%%%%%%%
% Acknowledgments
%%%%%%%%%%%%%%%%%%%%%%%%%%%%%%%%%%%%%%%%%%%%%%%%%%%%%%%%%%%%%%%%%%%%%%%%%%%%%%%
%
\section*{Acknowledgments}
We acknowledge the efforts of the engineers and technicians who have
participated in the construction and maintenance of L3 and express our
gratitude to the CERN accelerator divisions for the superb performance
of LEP.
%
%%%%%%%%%%%%%%%%%%%%%%%%%%%%%%%%%%%%%%%%%%%%%%%%%%%%%%%%%%%%%%%%%%%%%%%%%%%%%%%
% Bibliography
%%%%%%%%%%%%%%%%%%%%%%%%%%%%%%%%%%%%%%%%%%%%%%%%%%%%%%%%%%%%%%%%%%%%%%%%%%%%%%
%
% Style file to use with mcite.
% Use l3style with just cite.
\bibliographystyle{elsevier}

%\vfill
%These L3 Internal Notes are freely available upon request from \\
%The L3 Secretariat, CERN, CH--1211 Geneva 23, Switzerland. \\
%Internet: http://l3www.cern.ch.
%
%%%%%%%%%%%%%%%%%%%%%%%%%%%%%%%%%%%%%%%%%%%%%%%%%%%%%%%%%%%%%%%%%%%%%%%%%%%%%%
% Author List
%%%%%%%%%%%%%%%%%%%%%%%%%%%%%%%%%%%%%%%%%%%%%%%%%%%%%%%%%%%%%%%%%%%%%%%%%%%%%%
%
\newpage
\typeout{   }     
\typeout{Using author list for paper 178 -?}
\typeout{$Modified: Fri Jun 25 15:17:35 1999 by clare $}
\typeout{!!!!  This should only be used with document option a4p!!!!}
\typeout{   }
%
%
%
%  L A T E X  version!!
%
%
% Make sure that the Lep package has been used!
%\input{Lep.sty}%
%
%\ifx\LepCalled\undefined%
%\typeout{     }%
%\typeout{!!!!!!!!!!!!!!!!!!!!!!!!!!!!!!!!!!!!!!!!!!!!!!!!!!!!!!!!!!!}%
%\typeout{Yikes.  You haven't used the Lep package!}%
%\typeout{Please put \protect\usepackage\protect{Lep\protect} in your preamble,
%         followed by}%
%\typeout{\protect\Lep\protect{1\protect} or \protect\Lep\protect{2\protect}}%
%\typeout{     }%
%\typeout{For now you will get a Lep phase 2 authorlist (may not be right!).}%
%\typeout{!!!!!!!!!!!!!!!!!!!!!!!!!!!!!!!!!!!!!!!!!!!!!!!!!!!!!!!!!!!}%
%\typeout{     }%
%\Lep{2}\fi%

\newcount\tutecount  \tutecount=0
\def\tutenum#1{\global\advance\tutecount by 1 \xdef#1{\the\tutecount}}
\def\tute#1{$^{#1}$}
\tutenum\aachen            % 1
\tutenum\nikhef            % 2
\tutenum\mich              % 3
\tutenum\lapp              % 4
\tutenum\basel             % 5
\tutenum\lsu               % 6
\tutenum\beijing           % 7
\tutenum\berlin            % 8
\tutenum\bologna           % 9 
\tutenum\tata              % 10
\tutenum\ne                % 11
\tutenum\bucharest         % 12
\tutenum\budapest          % 13
\tutenum\mit               % 14 
\tutenum\florence          % 15
\tutenum\cern              % 16 
\tutenum\wl                % 17 
\tutenum\geneva            % 18
\tutenum\hefei             % 19
\tutenum\seft              % 20
\tutenum\lausanne          % 21
\tutenum\lecce             % 22
\tutenum\lyon              % 23
\tutenum\madrid            % 24
\tutenum\milan             % 25
\tutenum\moscow            % 26
\tutenum\naples            % 27
\tutenum\cyprus            % 28
\tutenum\nymegen           % 29
\tutenum\caltech           % 30
\tutenum\perugia           % 31
\tutenum\cmu               % 32
\tutenum\prince            % 33
\tutenum\rome              % 34
\tutenum\peters            % 35
\tutenum\salerno           % 36
\tutenum\ucsd              % 37
\tutenum\santiago          % 38
\tutenum\sofia             % 39
\tutenum\korea             % 40
\tutenum\alabama           % 41
\tutenum\utrecht           % 42
\tutenum\purdue            % 43
\tutenum\psinst            % 44
\tutenum\zeuthen           % 45
\tutenum\eth               % 46
\tutenum\hamburg           % 47
\tutenum\taiwan            % 48
\tutenum\tsinghua          % 49
{
\parskip=0pt
\noindent
{\bf The L3 Collaboration:}
\ifx\selectfont\undefined%  old style font selection
 \baselineskip=10.8pt
 \baselineskip\baselinestretch\baselineskip
 \normalbaselineskip\baselineskip
 \ixpt
\else%                      new style font selection
 \fontsize{9}{10.8pt}\selectfont
\fi
\medskip
\tolerance=10000
\hbadness=5000
\raggedright
\hsize=162truemm\hoffset=0mm
\def\r{\rlap,}
\noindent

M.Acciarri\r\tute\milan\
P.Achard\r\tute\geneva\ 
O.Adriani\r\tute{\florence}\ 
M.Aguilar-Benitez\r\tute\madrid\ 
J.Alcaraz\r\tute\madrid\ 
G.Alemanni\r\tute\lausanne\
J.Allaby\r\tute\cern\
A.Aloisio\r\tute\naples\ 
M.G.Alviggi\r\tute\naples\
G.Ambrosi\r\tute\geneva\
% ...
\newpage
%\rule{\textwidth}{0.4pt}
\begin{list}{A}{\itemsep=0pt plus 0pt minus 0pt\parsep=0pt plus 0pt minus 0pt
                \topsep=0pt plus 0pt minus 0pt}
\item[\aachen]
 I. Physikalisches Institut, RWTH, D-52056 Aachen, FRG$^{\S}$\\
 III. Physikalisches Institut, RWTH, D-52056 Aachen, FRG$^{\S}$
\item[\nikhef] National Institute for High Energy Physics, NIKHEF, 
     and University of Amsterdam, NL-1009 DB Amsterdam, The Netherlands
\item[\mich] University of Michigan, Ann Arbor, MI 48109, USA
\item[\lapp] Laboratoire d'Annecy-le-Vieux de Physique des Particules, 
     LAPP,IN2P3-CNRS, BP 110, F-74941 Annecy-le-Vieux CEDEX, France
\item[\basel] Institute of Physics, University of Basel, CH-4056 Basel,
     Switzerland
\item[\lsu] Louisiana State University, Baton Rouge, LA 70803, USA
\item[\beijing] Institute of High Energy Physics, IHEP, 
  100039 Beijing, China$^{\triangle}$ 
\item[\berlin] Humboldt University, D-10099 Berlin, FRG$^{\S}$
\item[\bologna] University of Bologna and INFN-Sezione di Bologna, 
     I-40126 Bologna, Italy
\item[\tata] Tata Institute of Fundamental Research, Bombay 400 005, India
\item[\ne] Northeastern University, Boston, MA 02115, USA
\item[\bucharest] Institute of Atomic Physics and University of Bucharest,
     R-76900 Bucharest, Romania
\item[\budapest] Central Research Institute for Physics of the 
     Hungarian Academy of Sciences, H-1525 Budapest 114, Hungary$^{\ddag}$
\item[\mit] Massachusetts Institute of Technology, Cambridge, MA 02139, USA
\item[\florence] INFN Sezione di Firenze and University of Florence, 
     I-50125 Florence, Italy
\item[\cern] European Laboratory for Particle Physics, CERN, 
     CH-1211 Geneva 23, Switzerland
\item[\wl] World Laboratory, FBLJA  Project, CH-1211 Geneva 23, Switzerland
\item[\geneva] University of Geneva, CH-1211 Geneva 4, Switzerland
\item[\hefei] Chinese University of Science and Technology, USTC,
      Hefei, Anhui 230 029, China$^{\triangle}$
\item[\seft] SEFT, Research Institute for High Energy Physics, P.O. Box 9,
      SF-00014 Helsinki, Finland
\item[\lausanne] University of Lausanne, CH-1015 Lausanne, Switzerland
\item[\lecce] INFN-Sezione di Lecce and Universit\'a Degli Studi di Lecce,
     I-73100 Lecce, Italy
\item[\lyon] Institut de Physique Nucl\'eaire de Lyon, 
     IN2P3-CNRS,Universit\'e Claude Bernard, 
     F-69622 Villeurbanne, France
\item[\madrid] Centro de Investigaciones Energ{\'e}ticas, 
     Medioambientales y Tecnolog{\'\i}cas, CIEMAT, E-28040 Madrid,
     Spain${\flat}$ 
\item[\milan] INFN-Sezione di Milano, I-20133 Milan, Italy
\item[\moscow] Institute of Theoretical and Experimental Physics, ITEP, 
     Moscow, Russia
\item[\naples] INFN-Sezione di Napoli and University of Naples, 
     I-80125 Naples, Italy
\item[\cyprus] Department of Natural Sciences, University of Cyprus,
     Nicosia, Cyprus
\item[\nymegen] University of Nijmegen and NIKHEF, 
     NL-6525 ED Nijmegen, The Netherlands
\item[\caltech] California Institute of Technology, Pasadena, CA 91125, USA
\item[\perugia] INFN-Sezione di Perugia and Universit\'a Degli 
     Studi di Perugia, I-06100 Perugia, Italy   
\item[\cmu] Carnegie Mellon University, Pittsburgh, PA 15213, USA
\item[\prince] Princeton University, Princeton, NJ 08544, USA
\item[\rome] INFN-Sezione di Roma and University of Rome, ``La Sapienza",
     I-00185 Rome, Italy
\item[\peters] Nuclear Physics Institute, St. Petersburg, Russia
\item[\salerno] University and INFN, Salerno, I-84100 Salerno, Italy
\item[\ucsd] University of California, San Diego, CA 92093, USA
\item[\santiago] Dept. de Fisica de Particulas Elementales, Univ. de Santiago,
     E-15706 Santiago de Compostela, Spain
\item[\sofia] Bulgarian Academy of Sciences, Central Lab.~of 
     Mechatronics and Instrumentation, BU-1113 Sofia, Bulgaria
\item[\korea] Center for High Energy Physics, Adv.~Inst.~of Sciences
     and Technology, 305-701 Taejon,~Republic~of~{Korea}
\item[\alabama] University of Alabama, Tuscaloosa, AL 35486, USA
\item[\utrecht] Utrecht University and NIKHEF, NL-3584 CB Utrecht, 
     The Netherlands
\item[\purdue] Purdue University, West Lafayette, IN 47907, USA
\item[\psinst] Paul Scherrer Institut, PSI, CH-5232 Villigen, Switzerland
\item[\zeuthen] DESY-Institut f\"ur Hochenergiephysik, D-15738 Zeuthen, 
     FRG
\item[\eth] Eidgen\"ossische Technische Hochschule, ETH Z\"urich,
     CH-8093 Z\"urich, Switzerland
\item[\hamburg] University of Hamburg, D-22761 Hamburg, FRG
\item[\taiwan] National Central University, Chung-Li, Taiwan, China
\item[\tsinghua] Department of Physics, National Tsing Hua University,
      Taiwan, China
\item[\S]  Supported by the German Bundesministerium 
        f\"ur Bildung, Wissenschaft, Forschung und Technologie
\item[\ddag] Supported by the Hungarian OTKA fund under contract
numbers T019181, F023259 and T024011.
\item[\P] Also supported by the Hungarian OTKA fund under contract
  numbers T22238 and T026178.
\item[$\flat$] Supported also by the Comisi\'on Interministerial de Ciencia y 
        Tecnolog{\'\i}a.
\item[$\sharp$] Also supported by CONICET and Universidad Nacional de La Plata,
        CC 67, 1900 La Plata, Argentina.
\item[$\natural$] Supported by Deutscher Akademischer Austauschdienst.
\item[$\diamondsuit$] Also supported by Panjab University, Chandigarh-160014, 
        India.
\item[$\triangle$] Supported by the National Natural Science
  Foundation of China.
\item[\dag] Deceased.
\end{list}
}
\vfill

%%% Local Variables: 
%%% mode: latex
%%% TeX-master: t
%%% End:

%%% Local Variables: 
%%% mode: latex
%%% TeX-master: t
%%% TeX-master: t
%%% TeX-master: t
%%% End: 

%%% Local Variables: 
%%% mode: latex
%%% TeX-master: t
%%% End: 

%%% Local Variables: 
%%% mode: latex
%%% TeX-master: t
%%% End: 

\newpage

\end{document}